# Nonlinear optical responses of semiconducting graphene nanoribbons: Strong constraint from a hidden symmetry

Jiawei Ruan[1,2,3], Weichen Tang[1,2], Yang-Hao Chan[4,5], Chen Hu[1,2], Xiaoxun Gong[1,2], and Steven G. Louie[1,2†]

[1] *Department of Physics, University of California at Berkeley, Berkeley, California 94720, USA*

[2] *Materials Sciences Division, Lawrence Berkeley National Laboratory, Berkeley, California 94720, USA.*

[3] *Eastern Institute of Technology, Ningbo 315200, China*

[4] *Physics Division, National Center of Theoretical Sciences, Taipei 10617, Taiwan*

[5] *Institute of Atomic and Molecular Sciences, Academia Sinica, Taipei 10617, Taiwan*

[†] sglouie@berkeley.edu

**Abstract:**

Nonlinear optical (NLO) responses are powerful tools for probing material structures, as they are highly sensitive to spatial symmetries. For instance, it is well known that having inversion symmetry forbids even-order NLO effects. Here, we show that nonspatial symmetries can also significantly constrain NLO responses. In graphene nanoribbons (GNRs) with bipartite-lattice structures, we identify the approximate bipartite-lattice particle-hole (BLPH) symmetry strongly suppresses even-order NLO effects, even for systems with no inversion symmetry. Importantly, we demonstrate that this approximate symmetry enables the optical generation of nearly pure spin shift currents in ferromagnetic GNRs, with the charge shift currents strongly suppressed. The spin orientation of these photo-induced DC spin currents can be easily tuned by an external magnetic field, providing distinct advantages over previous schemes for generating pure spin shift currents in antiferromagnetic systems. We illustrate our findings using a recently synthesized ferromagnetic Janus GNR. Our results underscore the significance of an overlooked symmetry aspect of NLO responses and highlight magnetic GNRs as promising candidates for spintronic devices.

**Main text:**

*Introduction.* Nonlinear optical (NLO) effects have garnered significant attention for decades due to their many fascinating physical phenomena and broad technological applications [1,2]. In particular, their sensitivity to crystal symmetry makes NLO responses exceptional probes for studying the structural properties of crystalline, nanostructured, and twisted moiré materials [3]. It is well established that in centrosymmetric materials, even-order optical responses, such as second-harmonic generation (SHG) and the bulk photovoltaic effect (BPVE), are strictly forbidden, resulting in the characteristic even-odd feature in high-harmonic generation (HHG) spectra [4]. The relationship between NLO susceptibility tensors and crystal symmetries has been thoroughly investigated [1]. However, the exploration of non-space-group symmetries that arise from internal properties within a system and their impact on NLO responses remains largely unexplored, with only limited cases considered so far [5].

Recent advances in bottom-up molecular precursor techniques have enabled the atomically precise design of a wide variety of graphene nanoribbons (GNRs), including control over their widths and diverse edge shapes [6–15]. These structurally precise 1D materials are of particular interest due to their tunable electronic and magnetic properties [16,17], making them promising candidates for nanoscale electronics devices [18]. Very recently, researchers have discovered a new class of GNRs—Janus graphene nanoribbons (JGNRs)—characterized by two distinct edge terminations. Notably, these JGNRs exhibit intrinsic ferromagnetism, offering new possibilities for spin-based applications [19].

A common characteristic of GNRs is their honeycomb structure composed of carbon atoms, with low-energy states predominantly derived from the $p_z$ orbitals. The next-nearest-neighbor (NNN) hopping energies in GNRs are typically very small, an order of magnitude smaller compared to nearest-neighbor hopping energy. For GNRs in structures with only six-fold rings, if the NNN hoppings are neglected, the system can be regarded as in a bipartite lattice, in which the atomic sites are divided into two distinct sublattices, typically labeled A and B, such that each site in sublattice A is only interacting with sites in sublattice B, and vice versa. Such systems within the tight-binding formalism exhibit additional symmetries not explicitly given by the crystal space group, including chiral symmetry and particle-hole symmetry, with the latter being the combination of chiral and time-reversal (TR) symmetries [20]. Using chiral symmetry, a topological classification framework has been developed for the electronic structure of GNR systems [21].

In this work, we demonstrate that the particle-hole symmetry in a bipartite lattice system—referred to as bipartite-lattice particle-hole (BLPH) symmetry—strongly constrains NLO responses in semiconducting graphene nanoribbons. Based on bipartite-lattice models, we show that such symmetry prohibits *any* even-order NLO responses even in structures *without* inversion symmetry. For a real GNR with small NNN hoppings, the BLPH symmetry in the Bloch band states is slightly broken, but the system still strongly suppresses the even-number harmonic generation coefficients and gives rise to an alternating even-odd emission intensity in HHG spectra. Furthermore, we show that in magnetic GNRs, although the chiral and TR symmetries are individually broken, their product, which leads to BLPH symmetry, remains approximately preserved. Taking the recently synthesized JGNR as a prototype material, we predict that this approximate symmetry enables optical generation of nearly pure spin shift currents in ferromagnetic GNRs, which is rarely achievable in conventional ferromagnetic materials. Our results underscore the significance of an overlooked symmetry aspect of NLO responses and position magnetic GNRs as compelling candidates for opto-spintronic technologies.

*BLPH symmetry and its effect on optical responses*. We begin by analyzing BLPH symmetry in a nonmagnetic bipartite-lattice model, and the case with spin degrees of freedom and broken TR symmetry will be discussed later. The real-space Hamiltonian of an ideal bipartite lattice is described by $H = \sum_{\langle ij \rangle} t_{ij} |i\rangle\langle j|$, with $t_{ij}$ a real number and $i \in$A-sites and $j \in$B-sites or vice versa with $\langle ij \rangle$ denoting nearest-neighbor pairs in real space. Such Hamiltonian possesses both chiral symmetry and BLPH symmetry. Firstly, it satisfies $H = -\Gamma H \Gamma^{-1}$, where $\Gamma$ is the chiral operator whose matrix form is diagonal with matrix elements equal to 1 for the A-sites and $-1$ for the B-sites (see End Matter for details and also Ref. [21]). Secondly, assuming TR symmetry holds for the system, there exist particle-hole symmetry, whose operation is defined as $C = \Gamma T$ with $T = \mathcal{K}$ where $\mathcal{K}$ is the complex conjugation. The particle-hole symmetry leads to $H = -CHC^{-1}$ for the real-space Hamiltonian and $H(-\mathbf{k}) = -CH(\mathbf{k})C^{-1}$ for the $\mathbf{k}$-space Hamiltonian. Because of this symmetry, the eigenstates of the Hamiltonian satisfy $|n\mathbf{k}\rangle = C|\bar{n}\bar{\mathbf{k}}\rangle$ (see End Matter for details) with energies $\epsilon_{n\mathbf{k}} = -\epsilon_{\bar{n}\bar{\mathbf{k}}}$ [20]. Here, $\bar{\mathbf{k}} \coloneqq -\mathbf{k}$, $\bar{n} \coloneqq -n$ and we adopt the band-index convention that the conduction bands are labeled by increasing positive integers away from the charge-neutral energy and the valance bands are labeled by decreasing negative integers. The above properties impose specific relations between matrix elements $O_{nm\mathbf{k}}$ and $O_{\bar{n}\bar{m}\bar{\mathbf{k}}}$ for a given operator $\hat{O}$. We have derived such relations for the Berry connection $\boldsymbol{\mathcal{A}}_{mn\mathbf{k}} = i\langle m\mathbf{k}|\partial_{\mathbf{k}}|n\mathbf{k}\rangle$, energy difference $\epsilon_{mn\mathbf{k}} = \epsilon_{m\mathbf{k}} - \epsilon_{n\mathbf{k}}$, occupation number difference $f_{mn\mathbf{k}} = f_{m\mathbf{k}} - f_{n\mathbf{k}}$, and velocity matrix elements $\mathbf{v}_{mn\mathbf{k}} = \hbar^{-1}\delta_{nm}\partial_{\mathbf{k}}\epsilon_{m\mathbf{k}} + \hbar^{-1} i\boldsymbol{\mathcal{A}}_{mn\mathbf{k}}\epsilon_{mn\mathbf{k}}$; and they are listed in Table A1 in the End Matter.

Now we see how BLPH symmetry impact the optical responses. We make our analysis based on the quantum kinetic equation of the density matrix within the independent-particle approximation for simplicity. The inclusion of electron self-energy effects (such as those in the time-dependent adiabatic GW method [22]) does not changes our essential symmetry-based findings, which can be found in SM. In the Bloch basis, the equation of motion (EOM) for the density matrix is given by [23],

$$i\hbar \frac{\partial}{\partial t}\rho_{mn\mathbf{k}}(t) = \epsilon_{mn\mathbf{k}}\rho_{mn\mathbf{k}} - e\mathbf{E}(t) \cdot [\mathbf{r}, \rho\,(t)]_{mn\mathbf{k}} - i\hbar\gamma\,(\rho(t) - \rho_0)_{mn\mathbf{k}}\,. \tag{1}$$

Here, $\gamma$ is a relaxation rate. $\rho_0$ is the equilibrium density matrix. The length gauge is adopted for the light-matter interaction, and the commutator relation with respect to the position operator is evaluated through the relation $[\mathbf{r}, \rho]_{mn\mathbf{k}} = [\boldsymbol{\mathcal{A}}, \rho]_{mn\mathbf{k}} + i\partial_{\mathbf{k}}\rho_{mn\mathbf{k}}$ with $\boldsymbol{\mathcal{A}}_{mn\mathbf{k}}$ being Berry connections defined above [24]. In the presence of an external optical field, the density matrix can be expanded in powers of the electric field $\mathbf{E}$, $\rho(t) = \rho^{(0)}(t) + \rho^{(1)}(t) + \rho^{(2)}(t) + \cdots$, where $\rho^{(N)} \propto E^N$. Using the symmetry relations in the first four rows of Table A1 and solving Eq. 1 order by order, we derive that the density matrix is constrained by BLPH symmetry and satisfies $\rho^{(N)}_{mn\mathbf{k}} = (-1)^{N-1}\rho^{(N)}_{\bar{n}\bar{m}\bar{\mathbf{k}}}$, meaning that for even orders, the value of the change in the density matrix elements associated with opposite $\mathbf{k}$ and band indices are exactly negative of each other; while for odd orders, they are the same (see detailed proof in Supplementary Materials (SM)). The current density of the N-th order may be obtained in terms of the corresponding order of the density matrix as $\mathbf{J}^{(N)}(t) = e\mathrm{Tr}\left[\hat{\mathbf{v}}\,\rho^{(N)}\right]$. Given that $v_{mn\mathbf{k}} = v_{\bar{n}\bar{m}\bar{\mathbf{k}}}$ (see Table A1 in the End Matter), the even order of $\mathbf{J}^{(N)}$ always vanishes due to the cancelation effect by contributions from the states with opposite momentum and band indices, while for the odd-order, these contributions are additive. This phenomenon is independent of whether the system has inversion symmetry or not, and is schematically illustrated in Fig. 1.

*Nonmagnetic semiconducting GNRs.* We demonstrate the BLPH symmetry effect on the optical responses for a specific example: a real graphene nanoribbon with only six-fold rings but without any rotational, inversion, or mirror symmetry in the atomic plane, referred to as a fully asymmetric GNR (faGNR), as shown in Fig. 2a. We specifically choose such a structure to eliminate the effect of space-group symmetry. We consider a tight binding (TB) model $H = H^0 + H'$, with $H^0 = -t_1\sum_{\langle ij\rangle}|i\rangle\langle j|$ and $H' = -t_2\sum_{\langle\langle i,j\rangle\rangle}|i\rangle\langle j|$, where $\langle\rangle$ and $\langle\langle\rangle\rangle$ represent nearest and next nearest neighbor pairs, respectively, with $t_1$ and $t_2$ as the corresponding hopping amplitudes. The band structures are plotted for the ideal bipartite-lattice model (with $t_1 = 2.7$eV and $t_2 = 0$) and the perturbated bipartite-lattice model (with $t_1 = 2.7$eV and $t_2 = 0.1t_1$), as shown in Figs. 2b and 2e, respectively. These parameters were obtained by fitting to our DFT computed results for the structure in Fig. 2a. As shown in Fig. 2e (and compared to Fig. 2b), the band structure only deviates slightly from the exact PH symmetrical features, indicating the $H'$ can be regarded as a small perturbation for GNRs.

We now calculate numerically the SHG of our faGNR to demonstrate the constraint from the BLPH symmetry and the effect of $H'$. The SHG conductivity tensor is defined as the ratio between the current density and electric field strength to the second order, $\sigma^{\mu\nu\lambda}(2\omega;\omega,\omega) = J_\mu(2\omega)/E_\nu(\omega)E_\lambda(\omega)$, where $\mu$, $\nu$, and $\lambda$ are Cartesian directions. $\sigma(2\omega;\omega,\omega)$ can be obtained by solving Eq. 1 to the second order, and the detailed expression can be found in SM. For analysis, we explicitly divided $\sigma$ into two parts $\sigma_a$ and $\sigma_b$, corresponding to contributions from states with $\mathbf{k}\in[\Gamma, X]$ and $\mathbf{k}\in[\Gamma, -X]$, respectively. As shown in Fig. 2c, for the ideal bipartite faGNR, $\sigma_a$ and $\sigma_b$ are indeed exactly opposite, leading to vanishing total conductivity. For the perturbated bipartite faGNR, the cancellation is still strong (nearly complete) in the low-frequency region (Fig. 2f).

Next, we calculate high harmonic generation produced by an electric field pulse and compare the difference between even and odd orders, employing a real-time propagation simulation of the density matrix [22,25,26]. The form of the pulse is given by $E(t) = E_0 f(t)\sin(\Omega t)$, where $\Omega$ is the pulse frequency, $E_0$ is the peak electric-field amplitude, and $f(t)$ represents an envelope function for the shape of the pulse (see more details in SM). We set the pulse frequency as $\hbar\Omega = 0.25$ eV. The HHG intensity is proportional to $|\omega\,\mathbf{J}(\omega)|^2$, where $\mathbf{J}(\omega)$ is the Fourier transformation of time-dependent current $\mathbf{J}(t)$. For the ideal bipartite faGNR, the even-order harmonics indeed disappear as expected (Fig. 2d). For the perturbated bipartite faGNR, the even-order harmonics peaks are visible but are much smaller than the nearby odd harmonics (Fig. 2g), meaning that they are still strongly constrained by the approximate BLPH symmetry.

*Magnetic GNRs*. The above symmetry analysis can be extended to the magnetic semiconducting GNR. We take the specific JGNR in Ref. [19] as an example, as it has been synthesized recently. Figure 3a shows its atomic structure, for which the numbers of atoms on the two sublattices are different. According to Lieb's theorem [27], a difference in the number of sublattice sites in the unit cell $N = |N_A - N_B|$ introduces $N$ zero-energy flat bands, which would lead to a ferromagnetic ground state with total spin $S = N/2$ per unit cell when electron-electron Coulomb interaction is included [27], as sketched in Fig. 3b. The ferromagnetism in this system has been shown theoretically to arise from topological edge states localized on the side of the JGNR with the straight zigzag edge, and these states have been observed in scanning tunneling spectroscopy measurements [19].

We first use a model Hamiltonian to analyze the underlying symmetry for a magnetic GNR. Based on a tight-binding model with Hubbard interactions at the mean-field level, which is usually a good description for graphene systems [28], we write our model as $\hat{H} = \hat{H}^0 + \hat{H}'$, with

$$\hat{H}^0 = -t_1 \sum_{\langle i,j\rangle,\sigma} c_{i\sigma}^{\dagger} c_{j\sigma} + \sum_{i,\sigma,\sigma'} \lambda_i c_{i\sigma}^{\dagger} s_{\sigma\sigma'}^{z} c_{i\sigma'} \quad (2)$$

and

$$\hat{H}' = -t_2 \sum_{\langle\langle i,j\rangle\rangle,\sigma} c_{i\sigma}^{\dagger} c_{j\sigma} , \quad (3)$$

where $\sigma = \uparrow$ or $\downarrow$ and the spin orientation is along an arbitrary quantization axis since there is negligible spin-orbit coupling (SOC) in the carbon system. The $\lambda_i$ term comes from the mean-field treatment of the on-site Hubbard interaction (see SM for the detailed derivation and also Ref. [29]). For our JGNR, this model reproduces the band structure obtained from our DFT calculations very well, as shown in Fig. 3c.

We find that the particle-hole symmetry of the Bloch band states is again exact if we consider $\widehat{H}^0$ alone (not shown in Fig. 3), which arises from the combination of TR and chiral symmetries although they are individually broken here. The corresponding symmetry operation is $\tilde{C} = \Gamma T = \Gamma(is_y\mathcal{K})$, where the TR operation $T$ is given by $is_y\mathcal{K}$ for magnetic systems with $s_y$ being the y-component Pauli matrix for the electron spin. Similar to the nonmagnetic case, it is easy to find that the k-space Hamiltonian has the relation $H^0(-\mathbf{k}) = -\tilde{C}^{-1}H^0(\mathbf{k})\tilde{C}$. The eigenstates for $H^0(\mathbf{k})$ satisfies $\tilde{C}|n\mathbf{k}\uparrow\rangle = |\bar{n}\bar{\mathbf{k}}\downarrow\rangle$, and the energies are related by $\epsilon_{n\mathbf{k}\uparrow} = -\epsilon_{\bar{n}\bar{\mathbf{k}}\downarrow}$, where the electron spin is a good quantum number and is flipped by the $\tilde{C}$ operator. In a similar manner, this symmetry imposes constraints between the matrix elements $O_{nm\mathbf{k}\sigma}$ and $O_{\bar{n}\bar{m}\bar{\mathbf{k}}\bar{\sigma}}$ that are associated with opposite bands, k-vectors and spin indices, as summarized in Table A2. Since the perturbating term $H'$ is small, the above features are approximately preserved for the full system, which explains why the bands from the full Hamiltonian in Fig. 3c are approximately symmetric about zero energy.

Now let's look at NLO responses in such magnetic systems. It is clear from Table A2 that for the even-order optical responses, the contributions from the two spin channels cancel each other under exact BLPH symmetry. For a real magnetic GNR with small BLPH symmetry breaking from next-nearest-neighbor hoppings, the cancellation at even orders is not strongly violated from our TB and DFT calculations. To demonstrate this, we calculate the shift current conductivity tensor $\sigma^{\mu\nu\lambda}_{\uparrow/\downarrow}$ of our ferromagnetic JGNR for the spin-up/spin-down channels (the expression of $\sigma^{\mu\nu\lambda}$ is given in the SM) based on the full DFT bands of our JGNR. Fig. 3d depicts the $zxz$ components (the $zxx$ and $zzz$ components vanish due to the mirror symmetry and are thus not plotted), which are responsible for the current along $z$ direction (along the length of the ribbon) under normal incident light with a polarization that has finite components along both the $x$ and $z$ axes. We see from the figure that, for a broad excitation frequency regime, the computed absolute value of the current density in the two spin channels are nearly identical, but the directions are opposite, indicating the charge current (proportional to $\sigma_{\text{charge}} = \sigma_\uparrow + \sigma_\downarrow$) is quenched but the spin current (proportional to $\sigma_{\text{spin}} = \sigma_\uparrow - \sigma_\downarrow$) is nonzero and promoted (see the inset of Fig. 3d). We note that this phenomenon remains largely preserved as the ribbon width increases, or when the JNGRs are placed on some insulating substrates, such as h-BN (see SM for more details).

We note that in the literature, the phenomenon of light-induced counter-propagating electrons with opposite spin polarization through a second-order NLO process is referred to as the spin bulk photovoltaic (SBPV) effect. This phenomenon has long been pursued for its potential to generate spin currents in an ultrafast timescale while minimizing the drawbacks associated with charge currents [30]. Previously, the SBPV effect has been proposed in antiferromagnetic (AFM) systems [31–33]. However, the resulting spin polarization of the spin current depends on the orientation of AFM Neel order, which is difficult to control. In our mechanism here, which arises from the BLPH symmetry, the photogeneration of nearly pure spin current can be achieved in a FM system. This makes the spin polarization of the spin current tunable by straightforwardly manipulating the direction of the magnetization of an FM sample with an external magnetic field (Fig. 3e). Moreover, by effectivity breaking the BLPH symmetry through mechanisms such as doping, the relative contributions of spin-up and spin-down currents can be controlled on demand (see SM). These features suggest that JGNRs can be engineered to provide versatile spin sources for optospintronics and related applications.

*Conclusions.* We have discovered that the existence of an effective particle-hole symmetry in graphene nanoribbons with near bipartite lattice symmetry imposes formidable constraints on their NLO responses, strongly suppressing even-order NLO responses even in structures lacking inversion symmetry. Applying our theory to the recently synthesized Janus GNR, we predict that the approximate BLPH symmetry uniquely facilitates the optical generation of pure spin currents in a ferromagnetic system, with tunability via an external magnetic field. This finding offers new opportunities for high-performance optospintronics using magnetic GNRs. Moreover, the BLPH symmetry constraints on NLO responses are expected to apply broadly to other near bipartite lattice materials, regardless of their shape and dimensionality, such as the Lieb lattice and dice lattice [34], providing a fundamental and universal principle governing the optical activity in these systems.

**Acknowledgments**

This work is supported by the National Science Foundation under grant DMR-2325410 which provided the symmetry analyses of matrix elements and nonlinear optical response formulation, and by the Center for Computational Study of Excited State Phenomena in Energy Materials (C2SEPEM) funded by the U.S. Department of Energy, Office of Science, Basic Energy Sciences, Materials Sciences and Engineering Division under Contract No. DE-AC02-05CH11231, as part of the Computational Materials Sciences Program which provided density functional calculations, density matrix calculations, and advanced codes. The work was also supported by the US Department of Energy, Office of Science, Basic Energy Sciences (BES), Materials Sciences and Engineering Division under contract DEAC02-05-CH11231 under the Nanomachine program (KC1203) which provided real-time density matrix simulations of HHG of faGNR. We acknowledge the use of computational resources at the National Energy Research Scientific Computing Center (NERSC), a DOE Office of Science User Facility supported by the Office of Science of the U.S. Department of Energy under Contract No. DE-AC02-05CH11231. The authors acknowledge the Texas Advanced Computing Center (TACC) at The University of Texas at Austin and the National Center for High-performance Computing (NCHC) for providing HPC resources that have contributed to the research results reported within this paper.

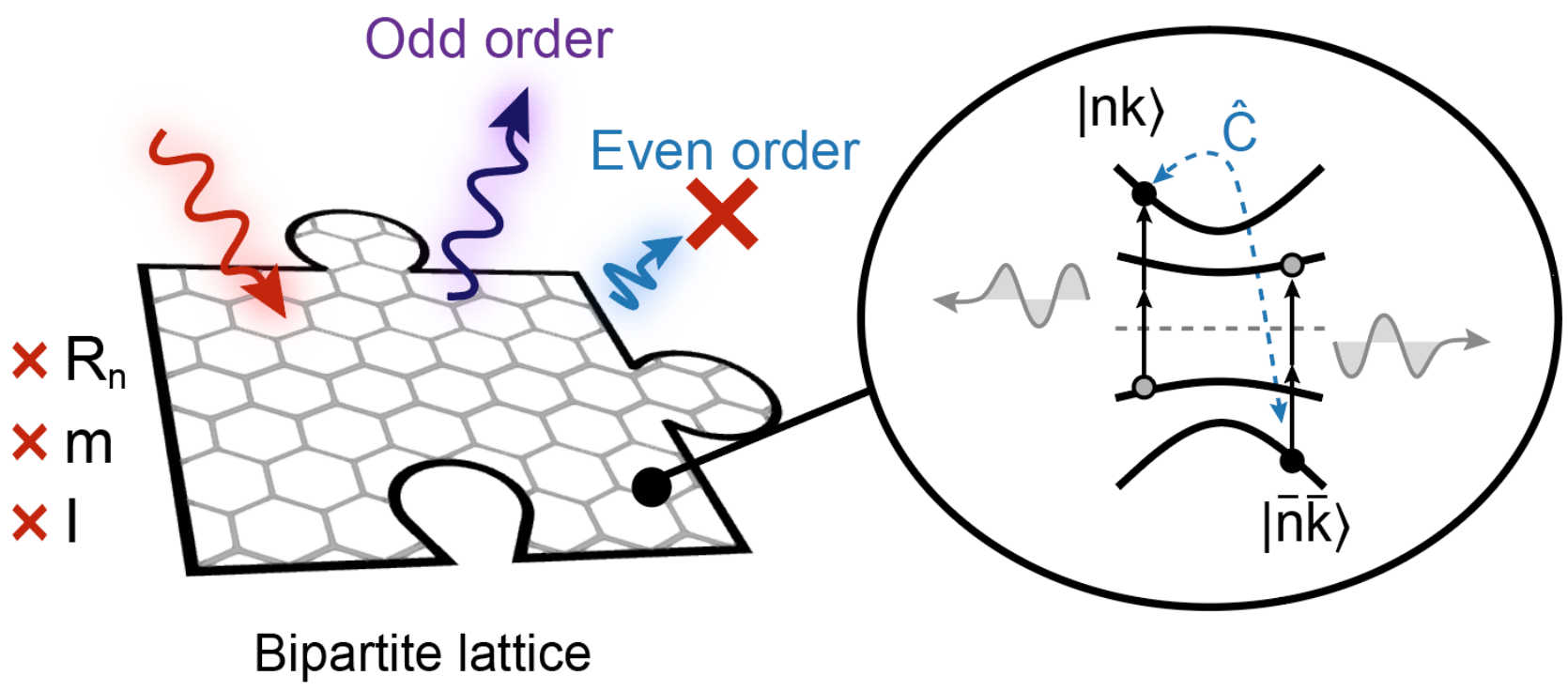

FIG. 1. Schematic of NLO responses in a bipartite-lattice semiconducting system. In such ideal system, there exist a nonspatial symmetry, particle-hole symmetry ($\hat{C}$), connecting states $|n\mathbf{k}\rangle$ and $|\bar{n}\bar{\mathbf{k}}\rangle$. This symmetry prohibits even-order NLO responses in any structure, even in the absence of any or all spatial symmetries such as rotational, mirror, or inversion symmetry.

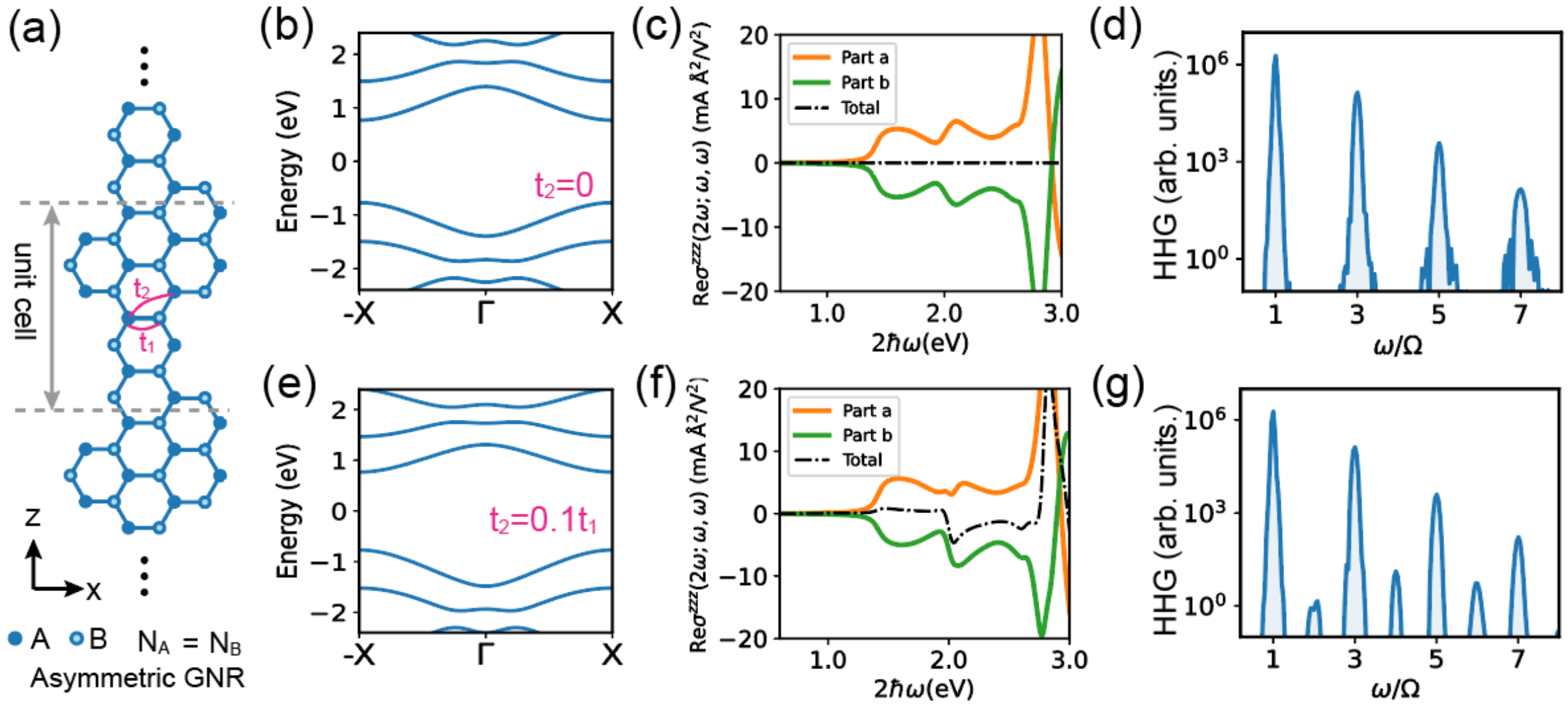

FIG. 2. (a) Crystal structure of a specific fully asymmetric GNR. There is no rotational symmetry, mirror symmetry, or inversion symmetry in the plane of the carbon atoms. (b) Band structure computed based on a TB model only considering nearest-neighbor hopping. (c) The real part of the frequency-resolved SHG conductivity tensor computed from the sum-over-band formula (Eq. S11 in the SM) with band structure from (b). The orange and green solid lines are the contributions from the states with $\mathbf{k} \in [\Gamma, X]$ (part a) and $\mathbf{k} \in [\Gamma, -X]$ (part b), respectively. (d) Higher harmonic generation spectra computed from real-time propagation of the density matrix with band structure from (b). A single pulse of duration 400fs with central frequency $\Omega = 0.25\text{eV}$ and electric field strength $E = 5\text{MV/cm}$ is used in the simulation. In this plot, the polarization of the pulse electric field and that of the emitted light are both taken along the $z$ direction. (e-g) are similar to (b-d) but are computed based on the TB model with next-nearest-neighbor hoppings included.

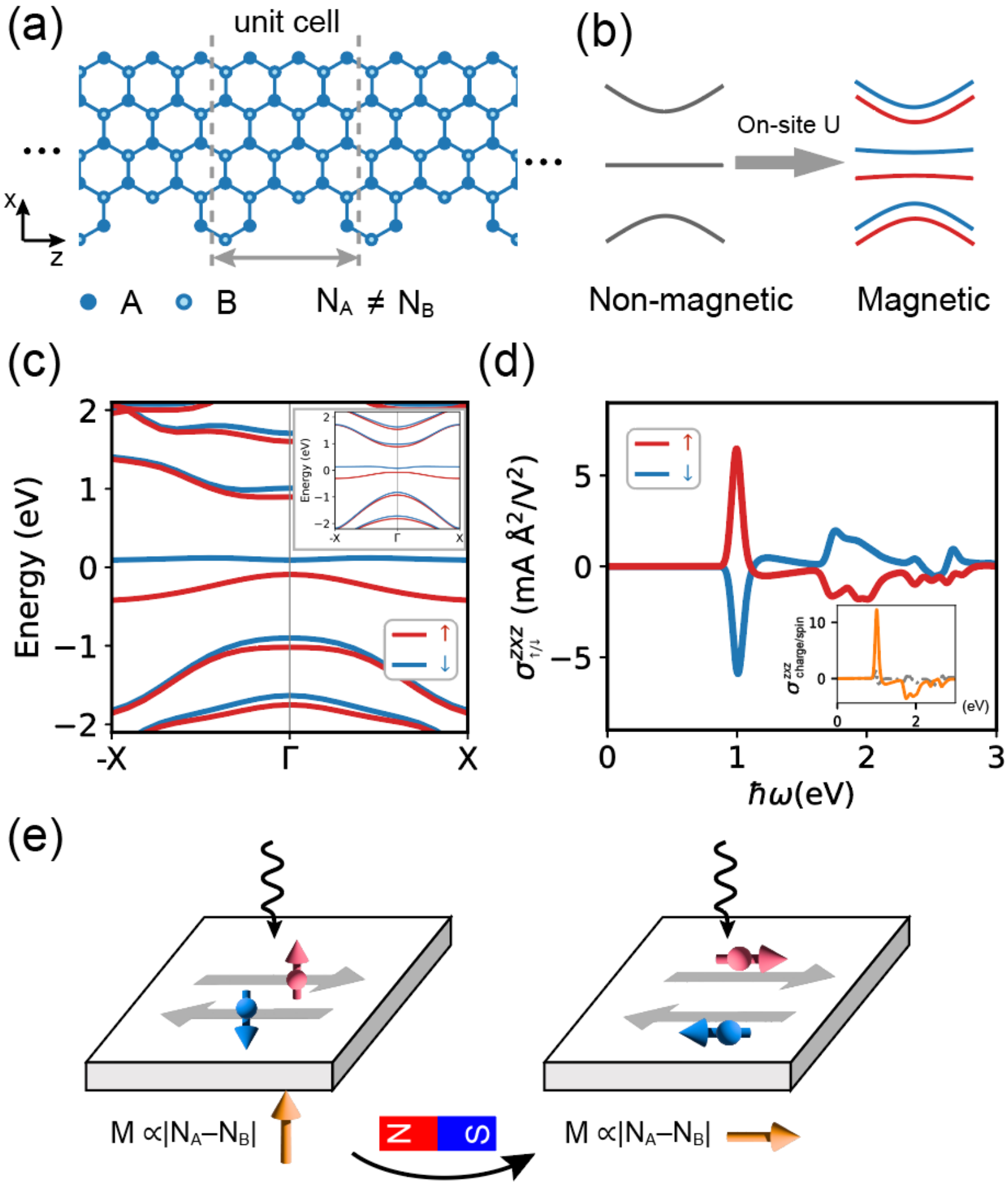

FIG. 3. (a) The structure of a Janus GNR, with different edge shapes on the two sides and different numbers of atoms on the graphene A and B sublattices, synthesized in Ref. [19]. In our DFT calculations, hydrogen atoms are used to passivate the dangling bonds (not shown); the corresponding structure is provided in the SM. (b) Schematic for the development of ferromagnetic electronic structure in a sublattice-imbalanced graphene nanoribbon after inclusion of on-site repulsive Coulomb interactions within a TB description. (c) Band structure of JGNR in (a) calculated by DFT within the LSDA approximation. Inset: Band structure calculated by TB model with mean-field Hubbard interaction. (d) Spin-resolved, frequency-dependent shift current conductivity tensor coefficients for spin ↑ and spin ↓ carriers. The inset shows that the charge shift current conductivity coefficient ($\sigma_{\mathrm{charge}} = \sigma_{\uparrow} + \sigma_{\downarrow}$, dashed gray) is largely suppressed compared to spin shift current conductivity coefficient ($\sigma_{\mathrm{spin}} = \sigma_{\uparrow} - \sigma_{\downarrow}$, orange solid). The y-axis in the inset uses the same units as the main figure. (e) Schematic depiction of the spin bulk photovoltaic effect realized in a FM system, illustrating that the spin orientation of the spin current is controllable with an external magnetic field.

## End Matter

*The representation of $\Gamma$ and $C$ operators in nonmagnetic bipartite lattices*. We first write out the components of wavefunctions $|n\mathbf{k}\rangle$ at a given energy $E$ explicitly as $\begin{pmatrix}\alpha_{n\mathbf{k}}\\ \beta_{n\mathbf{k}}\end{pmatrix}_E$, where $\alpha_{n\mathbf{k}}$ and $\beta_{n\mathbf{k}}$ are vectors respectively representing the A and B sublattice components of the state with band $n$ and wavevector $\mathbf{k}$. Then, for the chiral operator $\Gamma$, we have

$$\Gamma\begin{pmatrix}\alpha_{n\mathbf{k}}\\ \beta_{n\mathbf{k}}\end{pmatrix}_E=\begin{pmatrix}\alpha_{n\mathbf{k}}\\ -\beta_{n\mathbf{k}}\end{pmatrix}=\begin{pmatrix}\alpha_{\bar{n}\mathbf{k}}\\ \beta_{\bar{n}\mathbf{k}}\end{pmatrix}_{-E}, \tag{A1}$$

and for the particle-hole operator $C$, we have

$$C\begin{pmatrix}\alpha_{n\mathbf{k}}\\ \beta_{n\mathbf{k}}\end{pmatrix}_E=\begin{pmatrix}\alpha^*_{n\mathbf{k}}\\ -\beta^*_{n\mathbf{k}}\end{pmatrix}=\begin{pmatrix}\alpha_{\bar{n}\bar{\mathbf{k}}}\\ \beta_{\bar{n}\bar{\mathbf{k}}}\end{pmatrix}_{-E}, \tag{A2}$$

Here, symbols $E$ and $-E$ indicate the energies are symmetric with respect to the charge-neutral energy which is taken to be the zero energy.

*Constraints on matrix elements in nonmagnetic bipartite lattice from some nonspatial symmetries.* The form of Eq. (A2) will impose specific relations between matrix elements $O_{nm\mathbf{k}}$ and $O_{\bar{n}\bar{m}\bar{\mathbf{k}}}$ for a given operator $\hat{O}$. In Table A1, we list these relations for Berry connections $\boldsymbol{\mathcal{A}}_{mn\mathbf{k}}$, energy difference $\epsilon_{mn\mathbf{k}}$, occupation number difference $f_{mn\mathbf{k}}$, velocity matrix elements $\mathbf{v}_{mn\mathbf{k}}$ and N-th order of change of density matrix $\rho^{(N)}_{mn}(\mathbf{k})$. The relations imposed by the time-reversal and chiral symmetries are listed as well.

TABLE A1. Behavior of physical quantities of a spinless system with different discrete symmetries. Here $T$, $\Gamma$ and $C$ correspond to the time-reversal, chiral and particle-hole symmetry operations, respectively.

| | $T$ operation | $\Gamma$ operation | $C$ operation |
|---|---|---|---|
| $\mathcal{A}^a_{mn}(\mathbf{k})$ | $\mathcal{A}^{a*}_{mn}(-\mathbf{k})$ | $\mathcal{A}^{a*}_{\bar{n}\bar{m}}(\mathbf{k})$ | $\mathcal{A}^a_{\bar{n}\bar{m}}(-\mathbf{k})$ |
| $f_{mn}(\mathbf{k})$ | $f_{mn}(-\mathbf{k})$ | $f_{\bar{n}\bar{m}}(\mathbf{k})$ | $f_{\bar{n}\bar{m}}(-\mathbf{k})$ |
| $v^a_{mn}(\mathbf{k})$ | $-v^{a*}_{mn}(-\mathbf{k})$ | $-v^{a*}_{\bar{n}\bar{m}}(\mathbf{k})$ | $v^a_{\bar{n}\bar{m}}(-\mathbf{k})$ |
| $\epsilon_{mn}(\mathbf{k})$ | $\epsilon_{mn}(-\mathbf{k})$ | $\epsilon_{\bar{n}\bar{m}}(\mathbf{k})$ | $\epsilon_{\bar{n}\bar{m}}(-\mathbf{k})$ |
| $\rho^{(N)}_{mn}(\mathbf{k})$ | / | / | $(-1)^{N-1}\rho^{(N)}_{\bar{n}\bar{m}}(-\mathbf{k})$ |

*Constraints on matrix elements in magnetic systems in bipartite lattice from particle-hole symmetries*. In the SOC-free magnetic bipartite lattice system, the action of the particle-hole symmetry operator on the Bloch states satisfies $\tilde{C}|n\mathbf{k}\uparrow\rangle=|\bar{n}\bar{\mathbf{k}}\downarrow\rangle$, which can be expanded as

$$\tilde{C}\begin{pmatrix}\alpha_{n\mathbf{k}\uparrow}\\ \beta_{n\mathbf{k}\uparrow}\end{pmatrix}_E = \begin{pmatrix}\alpha^*_{n\mathbf{k}\uparrow}\\ -\beta^*_{n\mathbf{k}\uparrow}\end{pmatrix} = \begin{pmatrix}\alpha_{\bar{n}\bar{\mathbf{k}}\downarrow}\\ \beta_{\bar{n}\bar{\mathbf{k}}\downarrow}\end{pmatrix}_{-E}. \tag{A3}$$

Similarly to the nonmagnetic case, we can obtain specific relations between matrix elements $O_{nm\mathbf{k}\uparrow}$ and $O_{\bar{n}\bar{m}\bar{\mathbf{k}}\downarrow}$ for a given operator $\hat{O}$, which are listed in Table A2.

TABLE A2. Behavior of physical quantities of a magnetic system (which has negligible SOC) with bipartite-lattice particle-hole symmetry.

| | $\tilde{C}$ operation |
|---|---|
| $\mathcal{A}^a_{mn,\uparrow}(\mathbf{k})$ | $\mathcal{A}^a_{\bar{n}\bar{m},\downarrow}(-\mathbf{k})$ |
| $f_{mn,\uparrow}(\mathbf{k})$ | $f_{\bar{n}\bar{m},\downarrow}(-\mathbf{k})$ |
| $v^a_{mn,\uparrow}(\mathbf{k})$ | $v^a_{\bar{n}\bar{m},\downarrow}(-\mathbf{k})$ |
| $\epsilon_{mn,\uparrow}(\mathbf{k})$ | $\epsilon_{\bar{n}\bar{m},\downarrow}(-\mathbf{k})$ |
| $\rho^{(N)}_{mn,\uparrow}(\mathbf{k})$ | $(-1)^{N-1}\rho^{(N)}_{\bar{n}\bar{m},\downarrow}(-\mathbf{k})$ |